\documentclass{article}

\usepackage[final]{neurips_2019}

\usepackage[utf8]{inputenc} % allow utf-8 input
\usepackage[T1]{fontenc}    % use 8-bit T1 fonts
\usepackage{hyperref}       % hyperlinks
\usepackage{url}            % simple URL typesetting
\usepackage{booktabs}       % professional-quality tables
\usepackage{amsfonts}       % blackboard math symbols
\usepackage{nicefrac}       % compact symbols for 1/2, etc.
\usepackage{microtype}      % microtypography
\usepackage{graphicx}
\usepackage{bm}

\title{Detecting cutaneous basal cell carcinomas in 
ultra-high resolution and weakly labelled histopathological images}

\author{%
  Susanne Kimeswenger\\
  Department of Dermatology\\
  Kepler University Hospital \& \\
  Department of Experimental Physics \\
  Johannes Kepler University Linz\\
  Linz, Austria \\
  \texttt{susanne.kimeswenger@jku.at} 
   \And
  Elisabeth Rumetshofer,  Markus Hofmarcher\\
  Institute for Machine Learning\\
  Johannes Kepler University Linz\\
  Linz, Austria\\
  \texttt{rumetshofer@ml.jku.at} \\
  \texttt{hofmarcher@ml.jku.at} 
  \And
  \\
  \And
  Philipp Tschandl, Harald Kittler\\
  Department of Dermatology\\
  Medical University of Vienna \\
  \texttt{philipp.tschandl@meduniwien.ac.at}\\
   \texttt{harald.kittler@meduniwien.ac.at} 
  \And
  Sepp Hochreiter\\
  LIT AI Lab\\
  Johannes Kepler University Linz\\
  Linz, Austria\\
  \texttt{hochreit@ml.jku.at} 
  \And
  Wolfram Hötzenecker\\
  Department of Dermatology\\
  Kepler University Hospital Linz\\
  Linz, Austria\\
  \texttt{wolfram.hoetzenecker@kepleruniklinikum.at} 
  \And
  Günter Klambauer\\
  LIT AI Lab\\
  Johannes Kepler University Linz\\
  Linz, Austria\\
  \texttt{klambauer@ml.jku.at} 
}

\begin{document}

\maketitle

\begin{abstract}
  The diagnosis of basal cell carcinomas (BCC), one of the most common cutaneous malignancies in humans, is a   task regularly performed by pathologists and dermato-pathologists.
  Improving histological diagnosis by providing diagnosis suggestions, i.e. computer-assisted diagnoses, is a highly active area of research to improve safety, quality and efficiency.
  Machine learning methods are applied increasingly due to their superior performance. 
  However, typical images obtained by scanning histological sections often have a resolution that is
  not suitable for current state-of-the-art neural networks. Furthermore, network training is complicated by weak labels, because only a tiny fraction of the image indicates the
  disease class, whereas a large fraction of the image is highly similar to the non-disease class.
  The aim of this study is to evaluate whether it is possible to detect basal cell carcinomas in histological 
  sections using attention-based deep learning models and to overcome the ultra-high resolution and 
  the weak labels of whole slide images. 
  We demonstrate that attention-based models can indeed yield
  almost perfect classification performance with an AUC of 0.99.
\end{abstract}

\section{Introduction}
Basal cell carcinomas (BCCs) represent one of the most common cutaneous malignancies in humans \citep{Chinem2011}. 
Because of their frequency, BCCs are diagnosed by pathologists and dermato-pathologists on a regular basis. 
Digital pathology improves and simplifies histological diagnoses with regard to safety, 
quality and efficiency \citep{Griffin2017}. Digital pathology improves diagnoses of pathologists by providing diagnostic support, namely computer-assisted diagnoses \citep{Komura2018machine}. 
Recently this is mainly achieved using machine learning methods. 
Such methods could assist physicians, particularly pathologists in finding new histological 
patterns for the diagnosis of diseases.

From a machine learning perspective, convolutional neural networks \citep{Lecun1998gradient,Krizhevsky2012imagenet} would 
be the established method to tackle this image classification task, also because CNNs have been successfully
applied to analyze biological and medical images. Examples include the detection of melanoma, performing on par with 
dermatologists \citep{Tschandl2019, Haenssle2018, Esteva2017} or 
the prediction of cardiovascular risk factors based on retinal fundus 
images \citep{Poplin2018prediction}.
However, the size of typical histological images obtained in a resolution appropriate 
to see cellular structures are not suitable for processing with current state-of-the-art CNN architectures.
Such images often have a resolution of 50,000$\times$100,000 pixels, while CNNs are typically applied to images with a 
maximal resolution of up to 4,096$\times$4,096 pixels \citep{Momeni2018deep}. 
Recent attempts in training CNNs on histopathologic images frequently avoid the ultra-high resolution problem by sampling
random image patches from the full image \citep{cruzroa2013deep,Albarqouni2016aggnet,Janowczyk2016deep}, which
maintains the weak labelling problem. An overview of machine learning and deep learning 
methods that tackle the high-resolution problem of histopathology slides is given in \citet{Komura2018machine}. 
Furthermore, the classification of histopathology slides represents a weak label problem: the whole image slide is 
labelled with a single class (i.e. diagnosis), but a large fraction of the image is identical in all classes, 
and only a small region is indicative of the respective class.
Recently, multiple instance learning (MIL) and attention-based models have been proposed for analyzing whole slide 
images \citep{Tomczak2018histopathological,Ilse2018}. 

The aim of the study is to assess whether it is possible to detect basal cell carcinomas in whole slide images (WSI) overcoming the ultra-high resolution and weak label problems by adapting standard machine learning methods. 
Moreover, we aim to identify the \emph{key regions} in the image that are important for the decision of the  
predictive model and finally to compare them to the diagnostic key regions for board-certified pathologists.

\section{Massive Multiple Instance Learning with Attention}
\begin{figure}
    \centering
    \includegraphics[width=\textwidth]{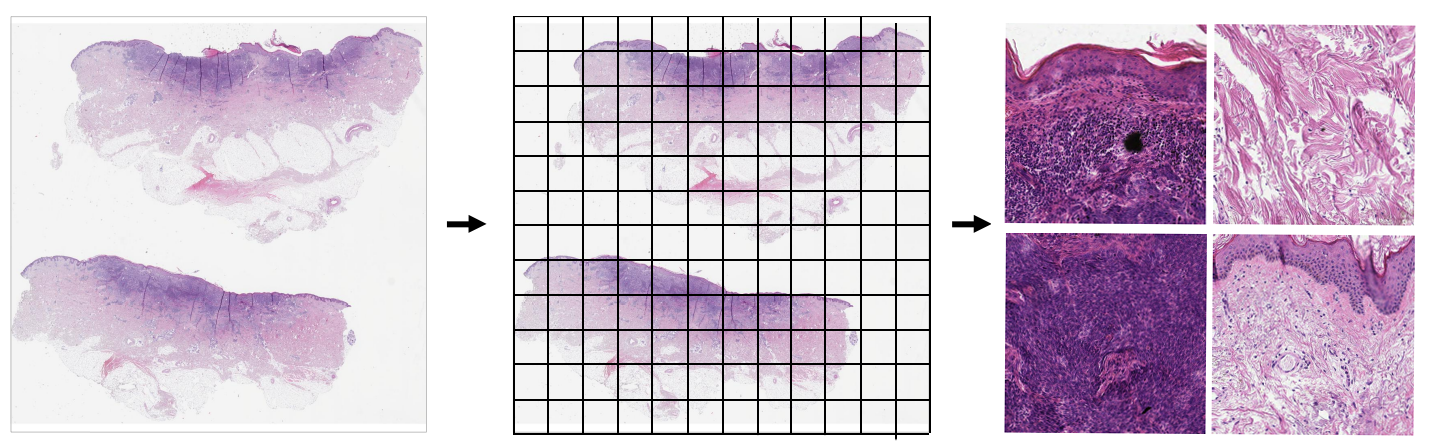}
    \caption{\textbf{ Left:} Histopathology slides with ultra-high resolution (usually >20,000 $\times$ >20,000 pixels) represent the input for the detection of BCC. 
             \textbf{ Center:} The full image is separated into patches with a resolution of 224 $\times$ 224 pixels. 
             \textbf{ Right:} Patches represent small regions of the histopathological slides and are used as instances in a multiple instance
             learning setting. 
             }
             \label{fig:tiling}
\end{figure}

We treat the problem of classifying extremely high resolution images as a multiple instance learning problem 
by dividing whole slide images into patches of relatively small resolution that can be processed by standard 
CNN architectures. This introduces the problem of credit assignment, i.e. how to combine signals from a massive 
number of patches to classify a full WSI as 'contains BCC' or 'does not contain BCC'. Simple solutions such as averaging 
patch predictions or adopting the prediction of the maximally activated patch have obvious problems with credit assignment. 
Therefore, we employ attention-based MIL pooling introduced as by Ilse and colleagues \citep{Ilse2018} and compare it against 
a baseline of downscaled WSI as well as patch based methods using mean and max pooling. An overview of the 
patch based approach is shown in Figure \ref{fig:tiling}. 
We use the well-known VGG11 architecture by \citet{Simonyan2014} for all patch-based experiments.

{\bf Baseline method.} 
As a baseline method, we use a standard CNN trained on whole slide images 
down-scaled to 1024$\times$1024 pixels. The down-sampling strategy means that only $\tilde{}$ $3\%$ of the available 
information was used for the classification of the histological sections.
The CNN consists of five blocks of convolution-convolution-maxpooling and utilizes SeLU activation functions~\citep{Klambauer2017}. 
The architecture and the hyperparameters of this CNN were optimized on a validation set using manual hyperparameter tuning.
The network was trained with standard SGD. Mean and standard deviation of accuracy, F1 score and the area under ROC curve (AUC) were calculated by re-training the networks 100 times. 

{\bf WSI method.}
All WSI were divided into patches of 224$\times$224 pixels covering the whole image, where padding was used if necessary. 
To exclude empty background patches, the average color intensity $c_p$ of all pixels for each patch $p$
was calculated. Then, for each WSI the maximum $c_{\mathrm{max}}=\max_p c_p$ was calculated and
all patches with $c_p$ higher than $95\%$ of this maximum were removed and considered as empty.
Each patch was normalized to zero mean and unit variance; no stain-normalization was applied.

Each non-empty patch was processed by the CNN and the activations of the last layer were stored as representation
of the respective patch. These activations were then used as input for the final classification network, 
where all patch activations of a single WSI were passed to the network as a mini-batch.
For mean and max pooling MIL methods we calculated the mean or maximum of the resulting network predictions.
The attention classifier was designed according to \citet{Ilse2018}, where the representation of 
each instance $\bm h_k$ corresponds to the network activations of one image patch. We use a weighted 
average of instances, where the weights are the output of a neural network that directly predicts these attention weights.
A set of $K$ instances in a single whole slide image is represented by $H=\{{\bm h_1,\ldots, \bm h_K}\}$, 
the MIL pooling function is ${\bm z} = \sum_{k=1}^{K}{a_k{\bm h}_k}$ and the the attention weight $a_k$ is 
\begin{equation} \label{eq:2}
a_k = \frac{\exp\{{\bm w}^\top \mathrm{tanh}({\bm V} \bm h_k^\top)\}}
{\sum_{j=1}^K{\exp\{{\bm w}^\top \mathrm{tanh}({\bm V} \bm h_j^\top)\}}}
\end{equation}
where ${\bm w} \in \mathbb{R}^{L\times 1}$ and ${\bm V} \in \mathbb{R}^{L\times M}$ are trainable parameters. 
Finally, a classification layer with sigmoid activation $\sigma$ 
is applied to provide the prediction $\hat y=\sigma(\bm v^\top \bm z)$. Again, we used standard SGD for training. Mean and standard deviation of accuracy, F1 score and AUC were calculated 100 times re-training the network.

\begin{figure}
    \centering
    \includegraphics[width=0.45\columnwidth]{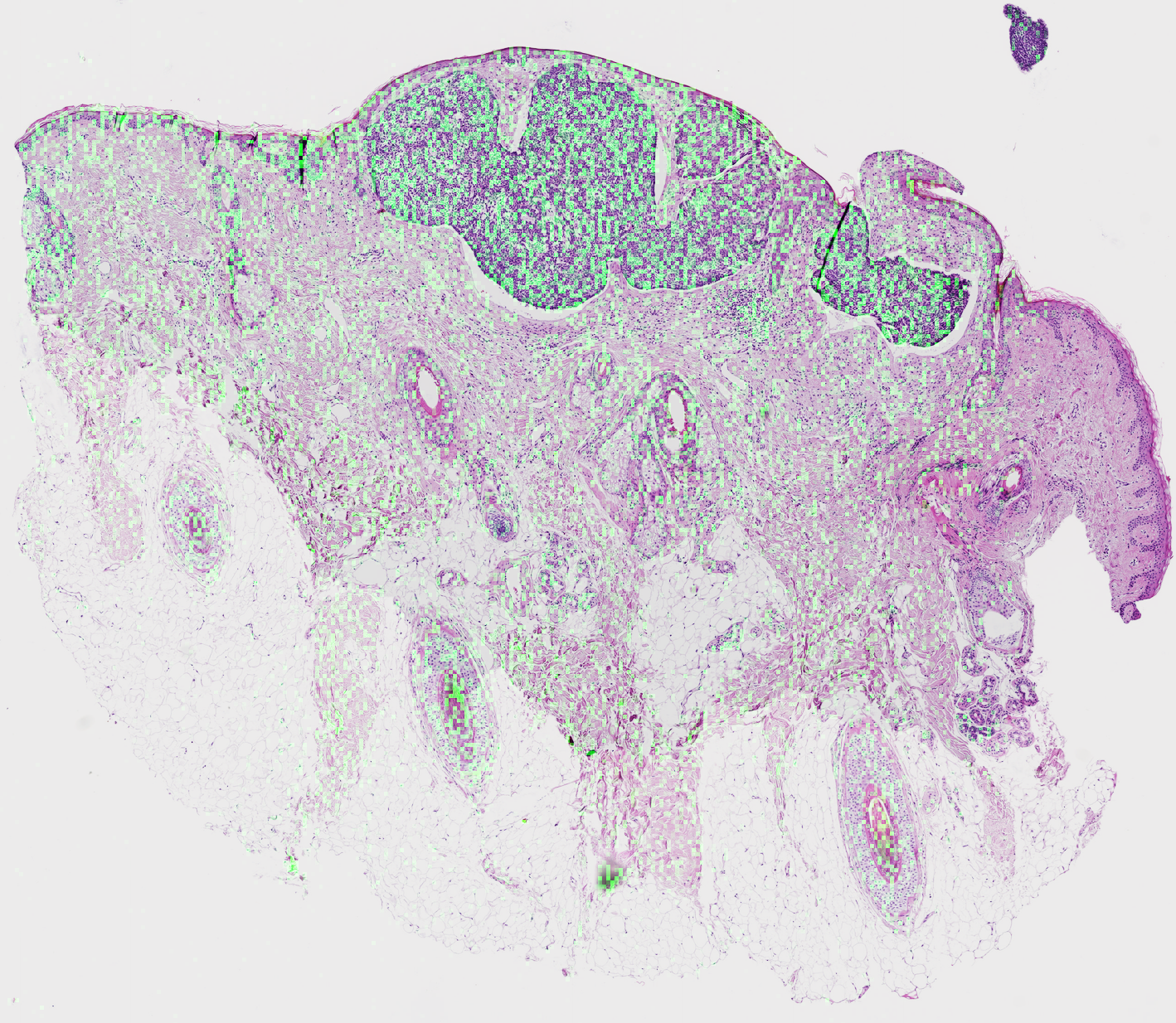}
    \includegraphics[width=0.45\columnwidth]{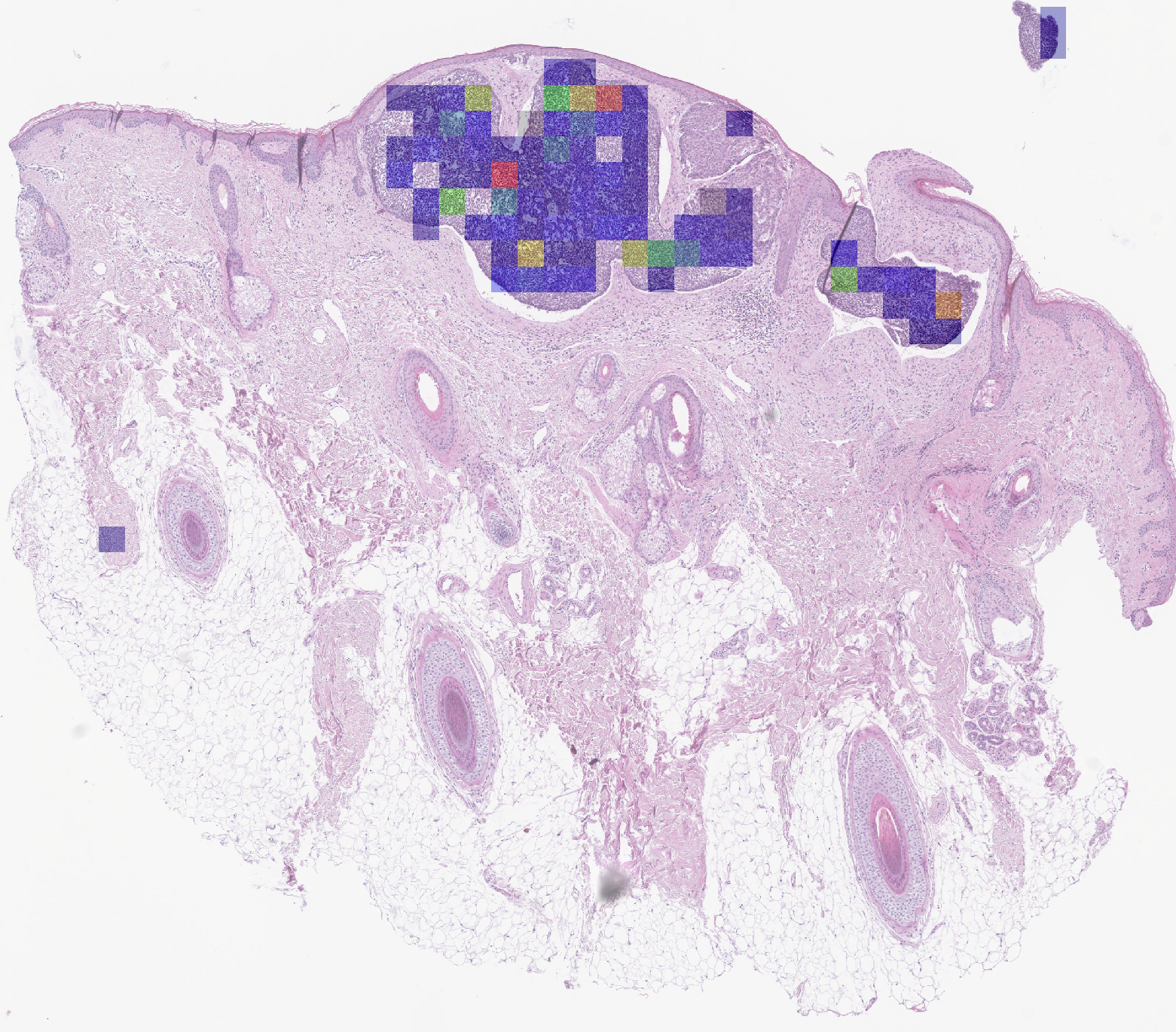}
    \caption{Visualization of the interpretation methods. \textbf{Left:} Regions
    of the input image contributing to the prediction according to the Integrated
    Gradients method are colored in green.
    \textbf{Right:} Regions
    of the input image contributing to the prediction according to the  
    weights provided by the attention mechanism.
    Both Integrated Gradients (left) and the attention mechanism (right) identify similar 
    regions as indicative of the cancer class.
    }
    \label{fig:interpretation}
\end{figure}

{\bf Interpretation method.} To interpret the predictions of the neural network, we used (1)
Integrated Gradients \citep{Sundararajan2017} for the baseline method and (2) the learned attention weights per image patch
of the attention mechanism described above. We visualized the attention weights in order to determine  the \emph{key regions} for the classification of the CNN model (Figure~\ref{fig:interpretation}). 

\section{Experiments and Results}
Images of histologicals slides of normal skin and skin sections containing BCCs were stained with hematoxylin and eosin (H/E, n=838 slides). 
647 of them represent BCC cases, 191 show normal skin. This set of 838 images 
was randomly split into 129 (15\%) test set and 709 (85\%) training images. 20\% of the training set
was used for as validation set.
The median size of the WSI is 56,896$\times$26,198 pixels, with the height ranging from 6,884 to 47,939
and the width ranging from 7,360 to 99,568 pixels. The images were retrospectively 
collected at the Medical University of Vienna and the Kepler University Hospital, according to ethics votes number 1119/2018 (Ethics Committee Upper Austria) and 2085/2018 (Ethics Committee Medical University of Vienna).

We trained the baseline method, and three methods based on mean pooling, max pooling, and the attention
mechanism described above using the PyTorch framework \citep{paszke2017automatic}. The results of the attention-based method against the baseline method and 
mean/max pooling MIL on our data set are shown in Table~\ref{tab:results}. The attention-based pooling method has significantly outperformed MIL with max pooling, MIL with mean pooling 
and an end-to-end CNN trained on down-sampled whole-slide images with respect to the AUC. 
Additionally, we report the accuracy and the F1 score of all predictive method in Table~\ref{tab:results}, which lead to the same ranking of the compared methods.
The sensitivity and the specificity of the MIL-attention method is .97 $\pm$ .01 and .91 $\pm$ .03, respectively. 
In addition to the classification, this attention-based architecture allows to easily identify the patches that are important 
for the classification of the network by inspecting the corresponding attention weights (see Figure~ \ref{fig:interpretation}). 

\begin{table}[]
    \begin{center}
        \caption{Performance metrics for the compared methods. Displayed values are the mean and the mean $\pm$ standard deviation across 100 re-runs of the training procedure. $p$-values correspond to Wilcoxon signed-rank test of AUC between CNN + MIL attention and the other methods. 
        \label{tab:results}}
        \resizebox{\textwidth}{!}{%
        \begin{tabular}{llcccl} 
        \toprule
        data type & method & accuracy  & F1 score & AUC  & $p$-value \\ 
        \midrule
        patches & CNN + MIL attention    & .96 (.95-.96) & .97 (.97-.98)  & \textbf{.99} (.99-.99) & \\ 
        patches & CNN + MIL max pooling  & .92 (.90-.94) & .95 (.94-.96)  & .98 (.98-.98) & <0.001\\ 
        patches & CNN + MIL mean pooling & .88 (.87-.90) & .93 (.92-.94)  & .93 (.93-.94)          & <0.001\\ 
        re-scaled WSI & end-to-end CNN   & .76 (.70-.82) & .83 (.77-.88)  & .90 (.89-.92)          &  <0.001\\ 
        \bottomrule
        \end{tabular}%
        }
        \end{center}
\end{table}

\section{Discussion}
Histopathology slides represent a gigantic source of information since they are collected 
and stored since decades. However, their computational analysis poses a huge challenge to 
machine learning techniques due to (1) the ultra-high resolution and (2) the weak labels. 
We have demonstrated that attention-based pooling and CNNs can be used to detect 
basal cell carcinomas in histopathology slides and how those models can be interpreted
and visualized.

{\bf Preliminary work.}
This extended abstract presents ongoing and preliminary work. We investigate other
types of attention mechanisms and experiments on additional data sets. 
Currently we use the VGG11 architecture for extracting features due to its simplicity.
We plan to compare this with other CNN architectures as well as to training the whole pipeline in an end-to-end fashion.
Furthermore, we want to scale down the WSI stepwise and find the optimal size to performance ratio. 
Another avenue we plan to follow is to compare the \emph{key regions} as identified via the attention weights
to saliency maps recorded via eye tracking of pathologists during diagnosis.

{\bf Acknowledgments.}
We would like to thank Rene Silye, Gudrun Lang, Giuliana Petronio and 
Christoph Sinz for their excellent scientific input and their great assistance in data 
collection. We thank the NVIDIA Corporation, Audi.JKU Deep Learning Center, Audi 
Electronic Venture GmbH, Janssen Pharmaceutica (MadeSMART), UCB S.A., FFG grant 871302, 
LIT grant DeepToxGen and AI-SNN, and FWF grant P 28660-N31.

% \section*{References}
%\section{References}
\bibliographystyle{apalike} 
\bibliography{references}

\end{document}